\documentclass[aps,twocolumn,amsmath,amssymb,showpacs,amsfonts,prd,nofootinbib]{revtex4}
\usepackage{epsfig}
\usepackage{graphicx}
\usepackage{dcolumn}
\usepackage{bm}
\usepackage{amsmath, amsthm, amssymb}
\usepackage{color}

\newcommand{\bea}{\begin{eqnarray}}
\newcommand{\eea}{\end{eqnarray}}

\begin{document}
\title{Holographic Screens and Transport Coefficients in the Fluid/Gravity Correspondence}
\author{Christopher Eling$^1$}
\author{Yaron Oz$^2$}
\affiliation{$^1$ SISSA, Via Bonomea 265, 34136 Trieste, Italy and INFN Sezione di Trieste, Via Valerio 2, 34127 Trieste, Italy}
\affiliation{$^2$ Raymond and Beverly Sackler School of
Physics and Astronomy, Tel-Aviv University, Tel-Aviv 69978, Israel}
\date{\today}
\begin{abstract}
We consider in the framework of the fluid/gravity correspondence the dynamics of hypersurfaces located in the holographic radial
direction at $r=r_0$. We prove that these hypersurfaces evolve, to all orders in the derivative expansion and including all higher curvature corrections, according to the same hydrodynamics equations with identical transport coefficients. The analysis is carried out for normal fluids as well as for superfluids.
Consequently, this proves the exactness of the bulk viscosity formula derived in  arXiv:1103.1657 via the null horizon dynamics.

\end{abstract}


\maketitle

The AdS/CFT correspondence \cite{Maldacena:1997re, Aharony:1999ti} relates certain strongly coupled large $N$ gauge theories to weakly coupled classical gravity, thus enabling the calculation of strongly coupled field theory quantities via General Relativity (GR).
In particular, the map of thermal states of field theories to classical gravity solutions containing black holes, relates the transport properties of gauge theory plasmas to black hole dynamics. There are two transport coefficients of relativistic uncharged hydrodynamics at the first viscous order: the shear viscosity $\eta$ and the bulk viscosity $\zeta$. If there are charges present in the system, there are additional conductivities at this order. The gravitational calculation of the ratio of the shear viscosity to the entropy density $s$ famously yields $\frac{\eta}{s} = \frac{1}{4\pi}$ \cite{Policastro:2001yc}.
Low values of this ratio are rather generic in strongly coupled gauge theories, and seem to characterize
the QCD plasma in heavy ion collisions.

In \cite{Eling:2011ms} the black hole dynamics was used to derive a simple formula for the ratio of the bulk viscosity to the shear viscosity:
\begin{align}
\frac{\zeta}{\eta} = \sum_i \left(s \frac{d \phi^{H}_i}{ds} + \rho^a \frac{d \phi^{H}_i}{d\rho^a} \right)^2 \ , \label{ratio}
\end{align}
where $\rho^a$ are the charge densities associated with the bulk gauge fields, and $\phi_i^H$ are the values of the bulk scalar fields on the horizon.
The derivation employs the null horizon focusing equation, and uses the equilibrium black hole solution.
When calculating the derivatives in (\ref{ratio}), one keeps all the parameters of the theory such as couplings and masses fixed. This, as demonstrated
in \cite{Buchel:2011yv,Buchel:2011wx}, may require in general the knowledge of the complete thermal equilibrium gravitational background, and not only the horizon data.

A natural question that arises is how transport coefficients, and in particular the bulk viscosity, depend on the holographic radial direction $r$ of the gravitational background, which in the field theory translates to a dependence on the renormalization group (RG) scale.
Transport coefficients depend on the couplings of the underlying microscopic field theory. Since the couplings in non-conformal field theories
depend generically on the RG scale, it follows that the transport coefficients also have this property.
Does this mean that
the formula (\ref{ratio}) has an approximate regime of validity, for instance only at high temperature?
There is evidence suggesting that this is not the case, and in fact it is the exact formula for the bulk viscosity \cite{Eling:2011ms,Buchel:2011yv,Buchel:2011wx,Patrushev:2011gm}. Still, a formal proof of its exactness is lacking.

In this letter we will prove a much stronger statement in the framework of the fluid/gravity  correspondence.
 We will study the dynamics of holographic screens (hypersurfaces) located in the radial direction at general $r=r_0$, where $r_0$ is a constant. We will prove that all these holographic screens evolve, to all orders in the derivative expansion including all higher curvature corrections, according to the same hydrodynamics equations with identical transport coefficients. The analysis will be carried out for normal fluids as well as for superfluids. An immediate consequence of this result is a proof of the exactness of  the bulk viscosity formula (\ref{ratio}).

We will consider  $(d+1)$-dimensional gravitational backgrounds holographically describing thermal states in  strongly coupled $d$-dimensional field theories. The $(d+1)$-dimensional gravitational action  is $I = I_{gravity}[g] + I_{matter}[g,\phi_i, A]$, where
\begin{align}
I_{gravity} = \frac{1}{16\pi} \int \sqrt{-g} d^{d+1} x \left(\Lambda + {\cal R}  + higher~curvature \right),
\label{grav}
\end{align}
where ${\cal R}$ is the Ricci scalar, $\Lambda$ a (negative) cosmological constant and ${\it higher~curvature}$  refers to general higher curvature terms.
We will set $\hbar = c = G^{d+1}_N=1$ throughout.

The matter action $I_{matter}[g,\phi_i, A]$ depends on (possibly charged) scalar fields $\phi_i$ and an (abelian) gauge field $A_{B}$.  We assume that it has the general form
\begin{align}
I_{matter} &= \frac{1}{16\pi} \int \sqrt{-g} d^{d+1} x \left( V_1(\phi_i \phi^{\star i}) F_{AB} F^{AB}  - \nonumber \right. \\ & \left. V_2(\phi_i \phi^{\star i}) \sum_i (D_A \phi_i) (\bar{D}^A \phi^{\star i}) + V_3(\phi_i \phi^{\star i})\right) \ ,
\label{matt}
\end{align}
where $F_{AB} = 2 \partial_{[A} A_{B]}$, $D_B = \nabla_B - i e A_B$ and $\bar{D}_B = \nabla_B + i e A_B$ for a given charge $e$.  The $V_i$ represent the potentials for the scalar fields.
The field equations of (\ref{matt}) have solutions that holographically describe non-conformal normal fluids as well as superfluids.

In this paper we will make use of two important conservation laws. The first is that
\begin{align}
E_{AB} \equiv \frac{2}{\sqrt{-g}}\frac{\delta I}{\delta g^{AB}}
\end{align}
satisfies
\begin{align}
\nabla^A E_{AB} = 0 \ . \label{bianchi}
\end{align}
This follows from the diffeomorphism invariance of the total action $I$ and imposing the field equations of the matter (scalars and gauge field). The second is that
\begin{align}
E^B \equiv \frac{2}{\sqrt{-g}}\frac{\delta I_{matter}}{\delta A_B}
\end{align}
satisfies
\begin{align}
\nabla_A E^A = 0 \ , \label{Maxwellidentity}
\end{align}
which follows from the gauge invariance of the matter action and imposing the scalar field equation.

On-shell, the metric field equations have the form $E_{AB} = 0$, or more specifically,
\begin{align}
E_{AB} =  G_{AB} + H_{AB} - T^{\rm{matt}}_{AB} = 0 \ .
\end{align}
$G_{AB}$ and $T^{\rm{matt}}_{AB}$ are the Einstein and matter stress-energy tensors respectively, while $H_{AB}$ represents any higher curvature corrections to the field equations. The field equations that arise from the variation of the matter action (\ref{matt}) are
\begin{align}
E^B = \nabla_A F^{AB} - J^B = 0 \ ,
\end{align}
where $J^A$ is a source current that exists when $e \neq 0$ and arises from the coupling between the scalars and the gauge field. The field equations of the scalars
will be unimportant to our discussion.

We will first consider solutions to this system that are dual to normal fluids on the boundary.  In this case we take the scalar fields to be neutral $e=0$ and real,
normalize the scalar and gauge field kinetic
terms by setting $V_1=1, V_2 = \frac{1}{2}$ and have $V_3$ an arbitrary potential for the scalar fields \cite{Eling:2011ms}.
In (radially shifted) Eddington-Finkelstein coordinates the metric solutions can be parametrized with the following metric ansatz
\begin{align}
g_{AB} dx^A dx^B  & = ds^2_{(0)} =  -f(r+R) u_\mu u_\nu dx^\mu dx^\nu  \nonumber\\
&+ 2 q(r+R) u_\mu dx^\mu dr +  h(r+R) P_{\mu \nu} dx^\mu dx^\nu \ . \label{zerothmetric}
\end{align}
Here we denote the coordinates of the bulk spacetime  by $x^A = (r, x^\mu), A=0,...,d$. The $x^\mu$ can be thought of as local coordinates on hypersurfaces of constant $r$. At $r=0$, the function $f$ vanishes and there is a horizon. The vector $u^\mu$ is a boost vector and has the dual role as a fluid velocity and the null normal to the horizon at $r=0$. The indices $\mu$, $\nu$ etc. are raised with the flat metric $\eta_{\mu \nu}$, and
$P_{\mu \nu} = \eta_{\mu \nu} + u_\mu u_\nu$
is the projection tensor. The entropy density associated with this solution is
$s= \frac{1}{4} h(R)^{\frac{d-1}{2}}$.

The function $q$ can be chosen to be a constant by a gauge choice corresponding to a redefinition of the radial coordinate $r$.
Thus, we expect that it will not play a role in the holographic description of the fluid.
Another way to see it, is to count the number of fields in the hydrodynamic description.
In the absence of charges we have the energy density $\varepsilon$, the pressure $p$ and the fluid velocity $u_{\mu}$, which is the same
number of actual fields in (\ref{zerothmetric}), namely $f$, $h$ and $u_{\mu}$.

In addition to the metric, there are the solutions to the scalar fields  $\phi_i(r+R)$ and the gauge field
\begin{align}
A_B dx^B = A(r+R) u_\mu dx^\mu \ .
\end{align}
The one function in the gauge field is in line with the one extra charge $\rho$ needed in the hydrodynamic description of a charged fluid.

Let us allow now the variables, along with $R$ and $u^\mu$ to be functions of $x^\mu$. The zeroth order metric \eqref{zerothmetric} and other fields as functions of $x^\mu$ are no longer solutions of the field equations and must be corrected order by order in a derivative expansion.
The fluid/gravity correspondence proposes that at the $n$th order in the derivative expansion \cite{Bhattacharyya:2008jc}, the bulk constraint field equations projected on the time-like boundary at spatial infinity ($r \rightarrow \infty$), whose unit space-like normal is $N_B$,
\begin{align}
E^{(n)}_{\mu B} N^B  = 0 \ ,
\end{align}
are the relativistic Navier-Stokes equations $\partial^\mu T^{\rm{fluid}}_{\mu \nu}  = 0$ at the $n$th viscous order. $T^{\rm{fluid}}_{\mu \nu}$ is the boundary stress-energy tensor. Similarly, in the presence of gauge fields, the Gauss law constraint projected on the time-like boundary at infinity
\begin{align}
N_A E^{(n) A} = 0 \ , \label{gauss}
\end{align}
is the additional current conservation law $\partial_\mu J^{\mu}_{\rm{fluid}} = 0$ in charged hydrodynamics, with $J^{\mu}_{\rm{fluid}}$ the boundary current \cite{Erdmenger:2008rm,Banerjee:2008th}. Alternatively, one can obtain these hydrodynamics equations from the Gauss-Codazzi equations of the horizon \cite{Eling:2009sj,Eling:2010hu}.

Consider now how the constraint equations depend on $r$, that is what happens when we project them on an arbitrary time-like hypersurface $r=r_0$. Suppose we have a gravitational solution at the $(n-1)$th order in derivatives, where $n \geq 1$, that is the bulk field equations are satisfied up to the $(n-1)$th order.
We would like to analyze the structure of $E_{AB}$ at $n$th order.
To impose the conservation law (\ref{bianchi}) at $n$th order we require that all the matter equations at this order satisfied. Then the conservation law takes the form
\begin{align}
\nabla^A_{(0)} E^{(n)}_{AB} = 0 \ ,
\end{align}
where the covariant derivative is with respect to the background zeroth order metric (\ref{zerothmetric}) and involves only radial derivatives.
The $\mu$ components of this equation read
\begin{align}
\nabla^A_{(0)} E^{(n)}_{A\mu} &= \frac{\partial}{\partial r} \left(-\frac{1}{q} E^{(n)}_{\mu \nu} u^\nu +
\frac{f}{q^2} E^{(n)}_{\mu r}\right) + \nonumber \\
& \frac{\partial}{\partial r} \ln( q h^{\frac{d-1}{2}}) \left(-\frac{1}{q} E^{(n)}_{\mu \nu} u^\nu + \frac{f}{q^2} E^{(n)}_{\mu r}\right) = 0 \ .
\label{bianc}
\end{align}
Using the unit normal to the hypersurface $r=r_0$,
$N^r = \frac{\sqrt{f}}{q},~N^{\mu} = -\frac{1}{\sqrt{q}}u^{\mu}$,
we can recast (\ref{bianchi}) as
\begin{align}
\frac{\partial}{\partial r} \left( \frac{\sqrt{f}}{q}  E^{(n)}_{\mu A}N^A\right) +
\frac{\partial}{\partial r} \ln( q h^{\frac{d-1}{2}} )\left( \frac{\sqrt{f}}{q}  E^{(n)}_{\mu A}N^A\right) = 0 \ .
\end{align}
Solving, we find that (off-shell) the constraints have the form
\begin{align}
E^{(n)}_{\mu A}N^A = \frac{F^{(n)}_\mu(x^\mu)}{h^{\frac{d-1}{2}} \sqrt{f}} \ ,
\label{constrfinal}
\end{align}
where $F^{(n)}_\mu(x^\mu)$ is some arbitrary function of $x^\mu$.

Taking the limit $r_0 \rightarrow \infty$, and renormalizing by an overall function of $r$, we get that $F^{(n)}_\nu(x^\mu) \sim \partial^\mu T^{\rm{fluid}}_{\mu \nu} = 0$ on-shell at the $n$th order. The factorization (\ref{constrfinal}) and the fact that the metric functions $f$ and $h$ are non-vanishing between the horizon and infinity lead to an important result: {\it the bulk constraint equations projected on any hypersurface $r=r_0$  are the same hydrodynamics equations with identical transport coefficients} \footnote{This general result has been verified explicitly in conformal hydrodynamics up to the first viscous order in \cite{Meyer}.}.

In the limit $r_0 \rightarrow 0$, $\sqrt{f}$ vanishes. At lowest orders in the derivative expansion, this is the near-horizon limit. It is well-known that in the limit as a ``stretched horizon" $r=r_0$ approaches the true horizon, $\sqrt{f} N^A$ becomes the horizon normal $u^\mu$  \cite{membrane}. Thus, we have
\begin{align}
\sqrt{f} E^{(n)}_{\mu A}N^A \rightarrow E^{(n)}_{\mu \nu} u^\nu \ .
\end{align}
In the case of Einstein gravity, the right hand side is the Gauss-Codazzi equations for the null hypersurface, one of which is the null focusing equation. Therefore, following the derivation of (\ref{ratio}) in \cite{Eling:2011ms}, we see that the ratio of the bulk viscosity to shear viscosity calculated at the horizon must be identical to the ratio that will be obtained from the boundary stress-energy tensor.  Remarkably, the horizon calculation of bulk viscosity from the focusing equation requires only the zeroth order solution, while for the computation of the boundary stress, one needs to have the bulk solution to first viscous order.
The bulk viscosity is an important transport coefficient of field theories. The exactness of the formula (\ref{ratio}) makes it a valuable tool in the study of strongly coupled non-conformal gauge theories and phenomenological holographic models of QCD.

A similar result for the current conservation equation in charged hydrodynamics can be obtained using the Gauss law constraint.
As before, assume that we have a solution to the gauge field equations $(n-1)$th order, so that the first non-zero piece of $E^A$ is at $n$th order. We must also impose the scalar field equations at $n$th order. Then the conservation law (\ref{Maxwellidentity}) reduces to
\begin{align}
\partial_r (q h^{\frac{d-1}{2}} E^{(n) r}) = 0 \ .
\end{align}
The constraint (\ref{gauss}) at $n$th order has the form
\begin{align}
N_A E^{(n) A} = \frac{q}{\sqrt{f}} E^{(n) r} \ .
\end{align}
Thus, we have an equation for the constraint, which yields
\begin{align}
N_A E^{(n) A} = \frac{G^{(n)}(x^\mu)}{h^{\frac{d-1}{2}} \sqrt{f}} \ ,
\end{align}
for some arbitrary function $G^{(n)}(x^\mu)$, which on-shell must be proportional to the boundary current conservation equations. The same analysis as above applies and this constraint is the same current conservation equation on each $r=r_0$ slice including the horizon, with identical transport coefficients.

Consider now solutions to the general gravitational system that holographically describe relativistic superfluids. In this case, the solutions can be parametrized with the following metric ansatz \cite{Sonner:2010yx,Bhattacharya:2011ee,Herzog:2011ec,Bhattacharya:2011tr}
\begin{align}
ds^2_{(0)} &=  -f(r+R) u_\mu u_\nu dx^\mu dx^\nu  \nonumber\\
&+ 2 q(r+R) u_\mu dx^\mu dr +  k(r+R) n_\mu n_\nu dx^\mu dx^\nu \nonumber\\
& +j(r+R) (u_\mu n_\nu + u_\nu n_\mu) dx^\mu dx^\nu + r^2 \tilde{P}_{\mu \nu} dx^\mu dx^\nu \ .
\end{align}
Here $n^\mu$ satisfies $u^\mu n_\mu = 0$, $n^\mu n_\mu = 1$ and $\tilde{P}_{\mu \nu} = \eta_{\mu \nu} + u_\mu u_\nu - n_\mu n_\nu$. The ansatz for the gauge field solution takes the form
\begin{align}
A_B dx^B = A(r+R) u_\mu dx^\mu + B(r+R) n_\mu dx^\mu \ .
\end{align}
The five functions in the metric and gauge field, together with $u^\mu$ and $n^\mu$, match the number of fields needed to characterize the
relativistic superfluid. In addition to the fields associated with the normal fluid $(\epsilon, p, \rho, u^\mu)$, we have a new vector field $n^\mu$, which is the component of the superfluid velocity orthogonal to the normal fluid velocity  $u^\mu$, the superfluid energy density $\epsilon_s$ and the superfluid
charge density $\rho_s$. Thus, as before, we expect that we can gauge away the function $q$.

We can now repeat the analysis of how the conservation laws effect the constraints for this ansatz. The result is a similar factorization of the constraints into a radial function times a function of $x^\mu$,
\begin{align}
E^{(n)}_{\mu A}N^A &= \frac{F^{(n)}_{s \mu}(x^\mu)}{r^{d-2} \sqrt{fk+j^2}} \ , \\
N_A E^{(n) A} &= \frac{G_s^{(n)}(x^\mu)}{r^{d-2} \sqrt{fk+j^2}} \ .
\end{align}
Therefore, we conclude that in the superfluid case there is again no running of the transport coefficients once the UV boundary conditions have been fixed.

It is straightforward to check that the inclusion of a Chern-Simons term in the bulk action, corresponding to having quantum anomalies in the dual field theory, does not change the above discussion. Thus, the parity violating anomalous transport coefficients in the hydrodynamics are independent of the radial direction, as expected from the field theory analysis for normal fluids \cite{Son:2009tf,Neiman:2010zi,KharzeevSecondOrder:2011ds,Loganayagam:2011mu} and superfluids
\cite{Bhattacharya:2011tr,Lin:2011mr,Neiman:2011mj}.

The above analysis applies also to the hydrodynamic expansion around Rindler space \cite{Eling:2009pb,Eling:2009sj}, where
the transport coefficients depend on the UV boundary conditions at the cutoff hypersurface $r=r_c$, but do not depend explicitly on the holographic radial direction $r$ from the UV to the IR \cite{Bredberg:2011jq}.

Finally, note that the study in this letter is different than the type of RG flow discussed in \cite{RG}.
In these works one imposes boundary conditions at a cutoff hypersurface and finds the corresponding (non-singular) interior bulk solution. Changing $r=r_c$ changes the solution and its corresponding energy-momentum tensor, leading to a RG flow in the transport coefficients. In our case, we consider the hypersurfaces $r_0$ in given solution.

\section*{Acknowledgements}

We would like to thank A. Buchel, U. Gursoy and A. Meyer for a valuable discussion.
The work is supported in part by the Israeli Science Foundation center
of excellence, by the US-Israel Binational Science Foundation (BSF),
and by the German-Israeli Foundation (GIF).


\begin{thebibliography}{References}

\bibitem{Maldacena:1997re}
J.~M.~Maldacena,
Adv.\ Theor.\ Math.\ Phys.\  {\bf 2}, 231 (1998)
[Int.\ J.\ Theor.\ Phys.\  {\bf 38}, 1113 (1999)]
[arXiv:h9ep-th/9711200].

\bibitem{Aharony:1999ti}
O.~Aharony, S.~S.~Gubser, J.~M.~Maldacena, H.~Ooguri and Y.~Oz,
Phys.\ Rept.\  {\bf 323}, 183 (2000)
[arXiv:hep-th/9905111].

\bibitem{Policastro:2001yc}
  G.~Policastro, D.~T.~Son, A.~O.~Starinets,
  Phys.\ Rev.\ Lett.\  {\bf 87}, 081601 (2001).
  [hep-th/0104066].

\bibitem{Eling:2011ms}
  C.~Eling, Y.~Oz,
  JHEP {\bf 1106}, 007 (2011).
  [arXiv:1103.1657 [hep-th]].

\bibitem{Buchel:2011yv}
  A.~Buchel,
  JHEP {\bf 1105}, 065 (2011).
  [arXiv:1103.3733 [hep-th]].

\bibitem{Buchel:2011wx}
  A.~Buchel, U.~Gursoy, E.~Kiritsis,
[arXiv:1104.2058 [hep-th]].

\bibitem{Patrushev:2011gm}
  A.~Patrushev,
  [arXiv:1107.2385 [hep-th]].

%
\bibitem{Bhattacharyya:2008jc}
S.~Bhattacharyya, V.~E.~Hubeny, S.~Minwalla and M.~Rangamani,
JHEP {\bf 0802}, 045 (2008)
[arXiv:0712.2456 [hep-th]].

\bibitem{Erdmenger:2008rm}
  J.~Erdmenger, M.~Haack, M.~Kaminski and A.~Yarom,
  JHEP {\bf 0901}, 055 (2009)
  [arXiv:0809.2488 [hep-th]].

\bibitem{Banerjee:2008th}
  N.~Banerjee, J.~Bhattacharya, S.~Bhattacharyya, S.~Dutta, R.~Loganayagam and P.~Surowka,
  arXiv:0809.2596 [hep-th].

\bibitem{Eling:2009sj}
C.~Eling and Y.~Oz,
  JHEP {\bf 1002}, 069 (2010)
  [arXiv:0906.4999 [hep-th]].
\bibitem{Eling:2010hu}
  C.~Eling, Y.~Neiman, Y.~Oz,
  JHEP {\bf 1012}, 086 (2010).
  [arXiv:1010.1290 [hep-th]].

\bibitem{Meyer}
 A.~Meyer,  M.Sc. Thesis, Tel-Aviv University (2011),
  arXiv:1107.0853 [hep-th].

\bibitem{membrane}
K. S. Thorne, R. H. Price and D. A. Macdonald, \textit{Black holes:
the membrane paradigm}, Yale University Press, 1986.

\bibitem{Sonner:2010yx}
  J.~Sonner and B.~Withers,
  Phys.\ Rev.\  D {\bf 82}, 026001 (2010)
  [arXiv:1004.2707 [hep-th]].

\bibitem{Bhattacharya:2011ee}
  J.~Bhattacharya, S.~Bhattacharyya, S.~Minwalla,
  JHEP {\bf 1104}, 125 (2011).
  [arXiv:1101.3332 [hep-th]].

\bibitem{Herzog:2011ec}
  C.~P.~Herzog, N.~Lisker, P.~Surowka and A.~Yarom,
  arXiv:1101.3330 [hep-th].

\bibitem{Bhattacharya:2011tr}
  J.~Bhattacharya, S.~Bhattacharyya, S.~Minwalla and A.~Yarom,
  arXiv:1105.3733 [hep-th].

\bibitem{Son:2009tf}
  D.~T.~Son and P.~Surowka,
  Phys.\ Rev.\ Lett.\  {\bf 103}, 191601 (2009)
  [arXiv:0906.5044 [hep-th]].

\bibitem{Neiman:2010zi}
  Y.~Neiman and Y.~Oz,
  JHEP {\bf 1103} (2011) 023
  [arXiv:1011.5107 [hep-th]].

\bibitem{KharzeevSecondOrder:2011ds}
  D.~E.~Kharzeev and H.~U.~Yee,
  arXiv:1105.6360 [hep-th].

\bibitem{Loganayagam:2011mu}
  R.~Loganayagam,
  arXiv:1106.0277 [hep-th].

\bibitem{Lin:2011mr}
  S.~Lin,
  arXiv:1104.5245 [hep-ph].

\bibitem{Neiman:2011mj}
  Y.~Neiman, Y.~Oz,
[arXiv:1106.3576 [hep-th]].

\bibitem{Eling:2009pb}
  C.~Eling, I.~Fouxon and Y.~Oz,
  Phys.\ Lett.\  B {\bf 680}, 496 (2009)
  [arXiv:0905.3638 [hep-th]].

\bibitem{Bredberg:2011jq}
  I.~Bredberg, C.~Keeler, V.~Lysov and A.~Strominger,
  arXiv:1101.2451 [hep-th];
  G.~Compere, P.~McFadden, K.~Skenderis and M.~Taylor,
  arXiv:1103.3022 [hep-th];
  G.~Chirco, C.~Eling and S.~Liberati,
  arXiv:1105.4482 [hep-th].

\bibitem{RG}
  I.~Bredberg, C.~Keeler, V.~Lysov and A.~Strominger,
  JHEP {\bf 1103}, 141 (2011)
  [arXiv:1006.1902 [hep-th]];
   S.~Kuperstein and A.~Mukhopadhyay, arXiv:1105.4530v2 [hep-th];
  D.~K.~Brattan, J.~Camps, R.~Loganayagam and M.~Rangamani,
  arXiv:1106.2577 [hep-th].



\end{thebibliography}
\end{document}